\documentclass[conference,a4paper]{APSIPA2020}
\usepackage{multirow}
\usepackage[dvips]{graphicx}
\usepackage{amsmath}
\usepackage[psamsfonts]{amssymb}
\usepackage{amsxtra}
\usepackage{threeparttable}

\usepackage{cite,balance}
\usepackage{epsfig,amssymb,amsmath,multirow}

\usepackage{color}
\usepackage{hyperref}
\usepackage{makecell}
\usepackage{balance}
\usepackage{amsmath}
\usepackage{xcolor,colortbl}

\hypersetup{
   unicode=false,          
   pdftoolbar=true,        
   pdfmenubar=true,        
   pdffitwindow=false,     
   pdfstartview={FitH},    
   pdftitle={My title},    
   pdfauthor={Author},     
   pdfsubject={Subject},   
   pdfcreator={Creator},   
   pdfproducer={Producer}, 
   pdfkeywords={keyword1} {key2} {key3}, 
   pdfnewwindow=true,      
   colorlinks=true,       
   linkcolor=blue,          
   citecolor=blue,        
   filecolor=magenta,      
   urlcolor=blue           
}

\begin{document}

\title{Classification of Speech with and without Face Mask using Acoustic Features}


\author{%
\authorblockN{%
Rohan Kumar Das and Haizhou Li
}
%
%
Department of Electrical and Computer Engineering,\\
National University of Singapore, Singapore\\
E-mail: \{rohankd, haizhou.li\}@nus.edu.sg
}

\maketitle
\thispagestyle{empty}

\begin{abstract}

The understanding and interpretation of speech can be affected by various external factors. The use of face masks is one such factors that can create obstruction to speech while communicating. This may lead to degradation of speech processing and affect humans perceptually. Knowing whether a speaker wears a mask may be useful for modeling speech for different applications. With this motivation, finding whether a speaker wears face mask from a given speech is included as a task in Computational Paralinguistics Evaluation (ComParE) 2020. We study novel acoustic features based on linear filterbanks, instantaneous phase and long-term information that can capture the artifacts for classification of speech with and without face mask. These acoustic features are used along with the state-of-the-art baselines of ComParE functionals, bag-of-audio-words, DeepSpectrum and auDeep features for ComParE 2020. The studies reveal the effectiveness of acoustic features, and their score level fusion with the ComParE 2020 baselines leads to an unweighted average recall of 73.50\% on the test set.

\end{abstract}

\section{Introduction}

Speech is a natural way of human-human and human-robot communication~\cite{HRI2012}. However, understanding and interpreting the conveyed message via speech gets affected by external factors, such as background noise and obstruction either at  speech source or at  receiver, that lead to performance degradation for various automatic systems as well as adverse effect in human perception.


There are various practical scenarios, where there is a requirement of wearing face masks. For instance, the surgeons working in operation theaters and forensic investigations are the most common among them. In addition, the current world situation due to COVID-19\footnote{www.who.int/emergencies/diseases/novel-coronavirus-2019} pandemic makes most of the people to wear face masks in their daily life. The wearing of face mask presents an adverse external factor to speech communication.

\footnotetext[2]{www.compare.openaudio.eu/}

The face mask detection has been performed previously to detect breach protocols in operating room using image processing techniques~\cite{Mask_maskless_2015}. However, such classification has never been investigated using speech~\cite{Schuller_ComParE2020}. Wearing of face mask may lead to increase in the vocal effort as face mask attenuates the speech energy, which is reported in a study conducted on oral examination data post-SARS~\cite{SARS_facemask}. Studies also show that spectral properties of fricatives are affected while using face masks~\cite{speaking_underCover}. We believe exploring different aspects of acoustic cues from a given speech can be helpful for detecting presence of face mask.

Literature shows that the surgical masks have less effect on speech understanding by human listeners~\cite{surgical_mask_2008}. A speech recognition study ~\cite{Mirco_surgeryRoom_2013} on data collected in surgery rooms  showed a high word error rate. Along similar direction, speaker recognition studies are also performed with different face masks~\cite{audio-visual_faceCover,facecover_is2015,Face_mask_is2016}. However, it is also found that the use of face masks does not have a large impact on speaker recognition performance~\cite{facecover_is2015,Face_mask_is2016}. Further, the identification of different face mask types showed that most of them could not be identified correctly~\cite{facecover_is2015}. 

The Computational Paralinguistics Evaluation (ComParE)\footnotemark[2] challenge series devotes on spearheading novel explorations for various paralinguistics studies~\cite{Schuller_IEEE_magazine}. It has been running successfully for more than a decade since its inception~\cite{SCHULLER2011speechComm}. The furtherance in the field of paralinguistics and computer science have advanced the state-of-the-art systems for various studies~\cite{SCHULLER2013_para,SCHULLER2019_lessons}. The latest ComParE 2020 runs three tasks, one of which is to find out whether a speaker is wearing a face mask for a given speech~\cite{Schuller_ComParE2020}. We report the participation of NUS team on this task of ComParE 2020 in this paper.



We consider three novel acoustic features capturing different acoustic properties of a signal. They are linear frequency cepstral coefficients (LFCC)~\cite{ASVspoof2015_lfcc}, instantaneous frequency cosine coefficients (IFCC)~\cite{Vijayan_Speech_comm} and constant-Q cepstral coefficients (CQCC)~\cite{CQCC_odyssey2016}. The LFCC captures spectral information using linearly spaced filterbanks, whereas IFCC captures the instantaneous phase of a signal. On the other hand, the CQCC features are derived using long-term constant-Q transform (CQT). All these three acoustic features have shown their effectiveness for different detection tasks previously~\cite{CQCC_odyssey2016, rkd_APSIPA2018,rkd_is2019_spoof,rkd_is2019_orca,rkd_ASRU2019}. We also consider widely popular mel frequency cesptral coefficient (MFCC)~\cite{Davis1980} feature as a contrast system and the ComParE 2020 baselines using ComParE functional feature set, bag-of-audio-words (BoAW), DeepSpectrum and auDeep feature~\cite{ComParE_functionals,openXBoW,DeepSpectrum,auDeep, audeep_fused}. Further, we perform a score level fusion of systems using acoustic features and the four ComParE 2020 baselines for the challenge submission.

The rest of the paper is organized as follows. Section~\ref{secii} describes the three acoustic features studied for finding presence of face masks. In Section~\ref{seciii}, the details of experiments are described. The results and analysis are reported in Section~\ref{seciv}. Finally, Section~\ref{conc} concludes the work.

\section{Acoustic Features}
\label{secii}

\begin{figure*}[t!]
\begin{center}
\includegraphics[width=1\textwidth]{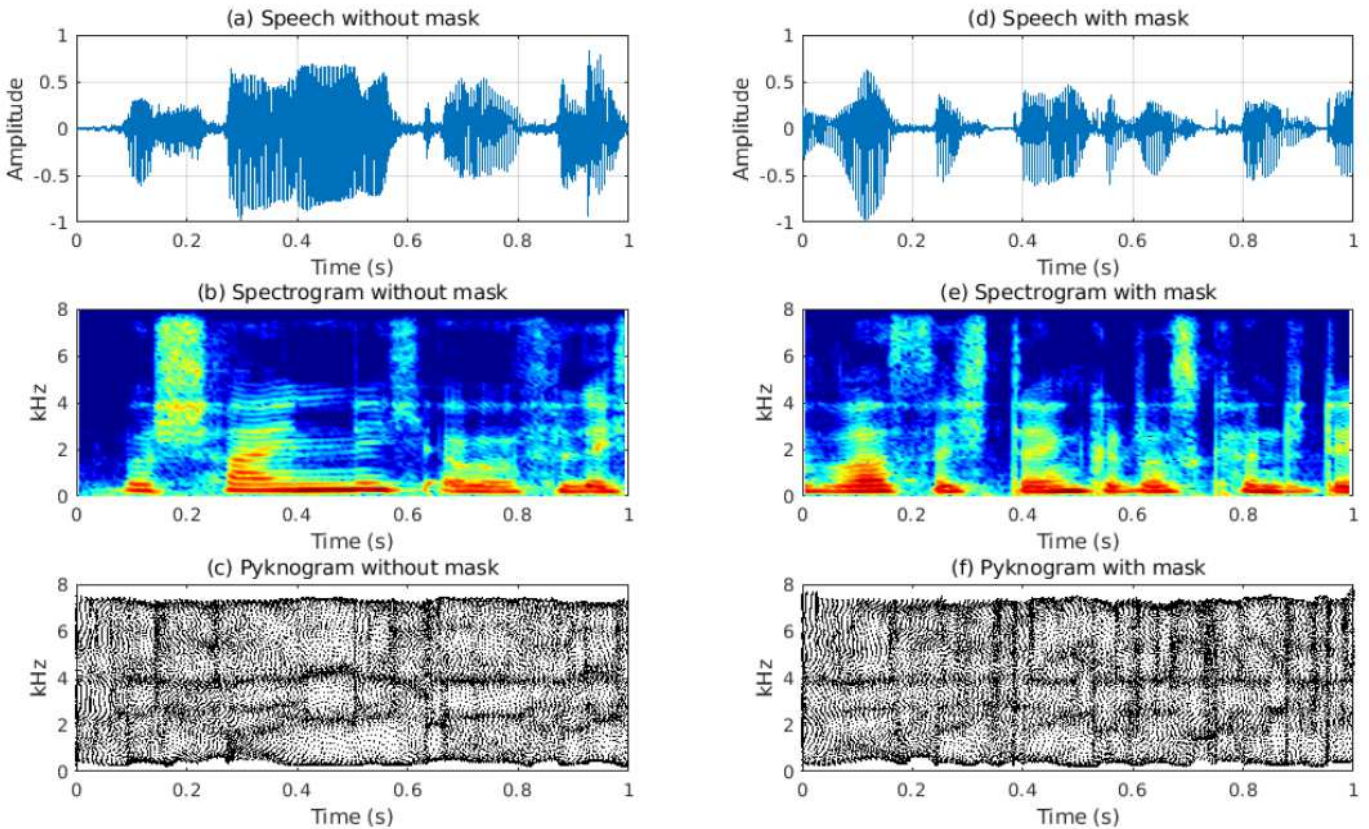}
\end{center}
\caption{(a) Speech without mask, its corresponding spectrogram and pyknogram in (b) and (c); (d) Speech with mask, its corresponding spectrogram and pyknogram in (e) and (f).}
\label{mask_plot}
\end{figure*}

In this section, we discuss three  acoustic features considered to capture the artifacts for classification of speech with and without masks. We next discuss each of them in detail. 

\subsection{Linear Frequency Cepstral Coefficients (LFCC)}

The short-term processing of speech signals followed by computation of log power spectrum is one of the most common ways of capturing acoustic artifacts from speech signal. The discrete cosine transform (DCT) over log power spectrum is taken to derive the cepstral coefficients in various speech processing application. Further, filterbanks are used to have a compact representation of the high dimensional cepstral features.

MFCC is one of the most widely used acoustic features that consider triangular filterbanks with a non-linear logarithmic mel scale, where the filters are placed densely in low frequency regions~\cite{Davis1980}. This is motivated by the human auditory perception~\cite{RevModPhys}. However, the same may not be applicable for machines to capture discriminative artifacts for classification or detection tasks as reported in~\cite{tifs_subband}. Therefore, we use linear filterbank based features. The LFCC feature replaces the mel filterbanks in MFCC by linearly spaced triangular filters~\cite{ASVspoof2015_lfcc}. As they focus on the artifacts uniformly along the frequency axis, they have been found to be useful for detection tasks previously~\cite{ASVspoof2015_lfcc}. Therefore, we consider LFCC as one of the feature to capture the acoustic properties from speech signal in this work.

\subsection{Instantaneous Frequency Cosine Coefficients (IFCC)}

Most of the acoustic features are derived using magnitude of the power spectrum. We consider that phase patterns of speech, in particular, aspirated plosives, are affected by the mask filter material. We would like to study the use of IFCC feature that is derived from analytic phase of a signal~\cite{Vijayan_Speech_comm}. The issue of phase warping is avoided by using Fourier transform properties to obtain the instantaneous frequency. The instantaneous frequency $\theta'$ for a signal in discrete-time $n$ can be derived as follows:

\begin{equation}
 \theta'[n]=\frac{2\pi}{N}Re\Bigg\{\frac{F_d^{-1}\{kZ[k]\}}{F_d^{-1}Z[k]}\Bigg\}
\end{equation}
where \({k=1, 2, \dots, K}\) represents the frequency bin index, $N$ is the length of the narrowband signal, $F_d^{-1}$ indicates inverse discrete Fourier transform and $Z[k]$ is the discrete Fourier transform of the analytic signal $z[n]$, obtained from the narrowband component of given signal~\cite{Vijayan_Speech_comm,Marple1997}.

The DCT is then applied on the instantaneous frequency components to obtain IFCC features\footnotemark[3]. These features carry long range acoustic information as short-term processing is performed only at the end to have frame-level features. As they are derived using the phase of a signal, they show complementary acoustic properties to many common features obtained from the magnitude spectrum. These features have been also successfully used for detection of spoofing attacks and orca activity~\cite{ASVspoof2017_Jelil2017,rkd_APSIPA2018,rkd_is2019_orca}.

\footnotetext[3]{https://ars.els-cdn.com/content/image/1-s2.0-S0167639316000364-mmc2.zip}
\footnotetext[4]{http://audio.eurecom.fr/software/CQCC\_v1.0.zip}

\subsection{Constant-Q Cepstral Coefficients (CQCC)}

We consider another aspect of acoustic property captured by long-term processing. The CQT is a long-term window transform~\cite{Brown1991} and is different from traditional features derived by short-term processing over a window of few milliseconds. It not only has a higher frequency resolution for lower frequencies, but also a higher temporal resolution for higher frequencies unlike the discrete Fourier transform. In addition, geometrically distributed center frequencies of each filter and the octaves make it unique, especially for detection of classification tasks~\cite{CQCC_odyssey2016,CQCC_CSL}. Previous studies used CQT to derive CQCC features that have been found effective for detection tasks~\cite{CQCC_odyssey2016,rkd_is2019_orca}. We believe they can also help to capture useful artifacts for classification of speech with and without face mask.

We note that uniform resampling applied to CQT based log power spectrum followed by DCT to derive the CQCC features\footnotemark[4]. For a given signal \({x(n)}\), its long-term transform CQT \({Y}(k,n)\) is computed as follows 
\begin{equation}	
 {Y}(k,n) %
  =\sum_{j=n-\lfloor{\frac{N_k}{2}}\rfloor}^{n+\lfloor{\frac{N_k}{2}\rfloor}}{x(j)}{a_k^\ast}{\Big(j-n-\frac{N_k}{2}\Big)}
 \label{eq:Qdefinition}	
 \end{equation}
where \({k=1, 2, \dots, K}\) represents frequency bin index, \({N_k}\) are the variable window lengths, \({a_k^\ast}(n)\) denotes the complex conjugate of \({a_k}(n)\), and \({\lfloor{\bullet}\rfloor}\) denotes rounding towards negative infinity. The basic functions \({a_k}(n)\) are complex-valued time-frequency atoms and are defined in~\cite{CQCC_odyssey2016}.

Fig.~\ref{mask_plot} shows a comparison of speech with and without mask along with their corresponding spectrogram and pyknogram. A pyknogram is a scatter plot denoting
the time-frequency representation of instantaneous frequencies from different filters in the filter-bank~\cite{pyknogram_ref}. We find the effect of having a mask is not that visible at the waveform level. However, it is observed that spectrograms of speech without mask has more prominent energy trajectories than that in the case of speech with mask. In addition, the pyknograms of speech with and without masks are different, showing phase information as another useful artifact. We believe the three features discussed above showing different acoustic properties can capture the attenuation in the frequency components of speech signal due to wearing of face masks. The information thus captured has definite potential to classify speech with and without mask.

\section{Experiments}
\label{seciii}

This section reports the experiments carried out for mask task of ComParE 2020. The database details and experimental setup are discussed in the following subsections.

\subsection{Database}
\label{seciii_1}

We used the Mask Augsburg Speech Corpus (MASC) released by organizers of ComParE 2020 for the studies~\cite{Schuller_ComParE2020}. The recordings are from 32 German native speakers that wear surgical mask from Lohmann and Rauscher. The corpus is gender balanced to have 16 male and 16 female speakers, whose age ranges from 20 to 41 years. The recordings are conducted in studio environment using large diaphragm condenser microphone. Although the original recordings are made in 48 kHz with 24 bit, the challenge participants are provided with 16 kHz mono/16 bit version. The duration of the corpus is around 10 hours. The data collection is made with and without wearing masks by the participants, where they answered some questions, read words known for their usage in medical operation rooms, drew a picture and talked about it, and described pictures. Further, the recordings are segmented into small duration non-overlapping 1-second segments for the challenge mask task to find whether a speaker is wearing mask or not.  

The collection of 1-second segments are then partitioned into train, development and test sets, which are released for the mask task of ComParE 2020. The labels are given for the segments of train and development set to indicate if the speech is recorded with or without face mask, whereas the test set is blinded for the challenge. Different explorations can be carried out using the train and development set to choose the best performing systems to apply on the test set. Table~\ref{table_database} presents a summary of the corpus released for mask task of ComParE 2020. The unweighted average recall (UAR) is used as metric for reporting the challenge results following the ComParE 2020 protocol~\cite{Schuller_ComParE2020}.

\begin{table} [t!]
\caption{\label{table_database} {A summary of the corpus for mask task of ComParE 2020.}}
\centerline{
\begin{tabular}{|c||c|c|c|c|}
\hline
{\bf Class} & {\bf Train} & {\bf Dev} & {\bf Test} & {\bf \# Utterances}\\
\hline
\hline
No-mask &  5,353 & 6,666 & blinded & blinded\\
Mask & 5,542 & 7,981 & blinded & blinded\\
\hline
\hline
Total & 10,895 & 14,647 & 11,012 & 36,554\\
\hline
\end{tabular}}
\vspace{-2mm}
\end{table}

\subsection{Experimental Setup}
\label{seciii_2}

ComParE 2020 organizers provide four state-of-the-art baseline systems for the mask task. Among these the ComParE functional feature based system is the official baseline similar to the previous editions~\cite{ComParE_functionals}. These features are obtained using the openSMILE\footnotemark[5] toolkit~\cite{Opensmile1,Opensmile2}. Another baseline with BoAW features is learned using low-level-descriptors (LLD) of ComParE feature set as well as the deltas of LDDs. The openXBoW\footnotemark[6] toolkit is used to extract the BoAW features for different codebook size~\cite{openXBoW}. The third baseline with DeepSpectrum features generated by DeepSpectrum\footnotemark[7] toolkit considers a pre-trained ResNet50 model~\cite{DeepSpectrum,cvpr16}. Lastly, unsupervised
representation learning with recurrent sequence to sequence autoencoders (S2SAE) are used to obtain the auDeep features using auDeep\footnotemark[8] toolkit that serves as the fourth baseline~\cite{auDeep,audeep_fused}. All the four baselines use support vector machine (SVM) as a classifier to classify speech with and without mask. The organizers also performed a majority voting based fusion among these baselines that serves as the benchmark system on the test set.

\footnotetext[5]{https://www.audeering.com/opensmile/}

\footnotetext[6]{https://github.com/openXBOW/openXBOW}

\footnotetext[7]{https://github.com/DeepSpectrum/DeepSpectrum}

\footnotetext[8]{https://github.com/auDeep/auDeep}

\footnotetext[9]{https://sites.google.com/site/bosaristoolkit/}

We now discuss the systems with acoustic features considered in this work. The LFCC and IFCC features are obtained for every frame of 20 ms with a shift of 10 ms. As the speech segments are of 1 second duration, we do not perform any voice activity detection. The parameters for CQT followed by CQCC feature extraction follows those given in~\cite{CQCC_odyssey2016}. We consider $\Delta$ and $\Delta\Delta$ coefficients for each acoustic feature along with the 30 static coefficients to have 90-dimensional feature representation. The widely used MFCC features in various speech processing applications are also considered as a contrast acoustic feature for the study. They are extracted with similar settings to that of LFCC, however, mel filterbanks are used in place of linear filterbanks.

Once the acoustic features are extracted, we use Gaussian mixture model (GMM) to build two different models of 512 mixtures each using features from speech with and without mask for each feature~\cite{Reynolds1995}. The choice of GMM for this task follows our previous work of orca activity detection~\cite{rkd_is2019_orca}. During testing, for a given test speech, its respective acoustic features are extracted and then likelihood is computed against the two models. Finally, the test speech is classified as the category showing a higher likelihood score. 

\begin{table} [t!]
\caption{\label{table_baseline} {ComParE 2020 baseline system results in UAR (\%) with ComParE Functionals, ComParE BoAW, DeepSpectrum and auDeep features for the mask task. A comparative analysis of the four systems with different parameters that include $C$: Complexity parameter of the SVM, $N$: Codebook size for BoAW, $X$: Power levels clipped below threshold. The four best baseline results are shown in bold face fonts.~\cite{Schuller_ComParE2020}}}
\centerline{
\begin{tabular}{|c|c|c|}
\hline
{\bf Parameters}& {\bf Development} & {\bf Test}\\
\hline
\hline
$C$ &\multicolumn{2}{|c|}{\bf ComParE Functionals + SVM}\\
\hline
$10^{-5}$ & 56.8 & 59.8\\
${10^{-4}}$ & {60.3} & {67.7}\\
$ {10^{-3}}$ & 62.3 & {67.8}\\
${\mathbf 10^{-2}}$ & {\bf 62.6} & 66.9\\
\hline
\hline
$N$ &\multicolumn{2}{|c|}{\bf ComParE BoAW + SVM}\\
\hline
125 & 59.8 & 58.7\\
250 & 61.5 & 62.7\\
500 & 63.1 & 65.0\\
1000 & 63.6 & 66.1\\
{\bf 2000} &{\bf 64.2} &{\bf 67.7}\\
\hline
\hline
{ Network} &\multicolumn{2}{|c|}{\bf DeepSpectrum + SVM}\\
\hline
{\bf ResNet50} & {\bf 63.4} & {\bf 70.8}\\
\hline
\hline
$X~dB$ &\multicolumn{2}{|c|}{\bf auDeep: S2SAE + SVM}\\
\hline
-30 & 60.1 &  57.4\\
-45 & 61.3 &  60.3\\
-60 & 61.9 &  61.6\\
-75 & 61.6 &  62.2\\
{\bf Fused} & {\bf 64.4} & {\bf 66.6}\\
\hline
\hline
{\bf Method} & \multicolumn{2}{|c|}{\bf Fusion: Benchmark Baseline}\\
\hline
Majority Vote 4 Best & - & {\bf 71.8}\\
\hline
\end{tabular}}
\end{table}

In general, the combination of multiple systems with complementary information helps to improve performance~\cite{rkd_is,rkd_ncc_2015,rkd_jasa}. Hence, we performed fusion of various acoustic feature based systems as well as the four baselines in this work. The fusion is carried out with logistic regression at score level using Bosaris\footnotemark[9] toolkit~\cite{Bosaris}, which applies the weights of various systems learned on the development set to the unseen test set.

\section{Results and Analysis}
\label{seciv}

We  re-implement the four baseline systems of ComParE 2020 mask task, and report their results in Table~\ref{table_baseline}. It is noted that the results on the test set are referred from ~\cite{Schuller_ComParE2020}. The best configuration of each of the baseline are chosen by tuning various parameters that are shown in bold fonts. The majority voting based fusion of these best baselines showing UAR of 71.8\% serves as the benchmark challenge performance.

\begin{table} [t!]
\caption{\label{table_proposed} {Performance comparison in UAR (\%) among the four acoustic feature systems, the majority voting fusion based benchmark baseline, the score level fusion of the acoustic features, 4 best baseline systems, and the all the systems.}}
\centerline{
\begin{tabular}{|c|c|c|}
\hline
{\bf Acoustic Feature}& {\bf Dev} & {\bf Test}\\
\hline
\hline
LFCC &  66.24  & 68.80 \\
IFCC & 60.31 & -\\
CQCC & 63.90 & -\\
MFCC & 61.52 & -\\
\hline
\hline
\multicolumn{3}{|c|}{\bf Benchmark: Majority Voting Fusion}\\
\hline
4 Best Baselines~\cite{Schuller_ComParE2020} & - & 71.80\\
\hline
\hline
\multicolumn{3}{|c|}{\bf Score Level Fusion}\\
\hline
4 Best Baselines & 68.38 &  72.10\\
4 Acoustic Features &{67.43} & 70.60\\
\hline
\hline
All & {\bf 69.84} &{\bf 73.50}\\
\hline
\end{tabular}}
\vspace{-2mm}
\end{table}

We now focus on the acoustic feature based systems and their fusion with other systems. It is noted that results on the test set are not available for all the systems as ComParE 2020 allows only 5 score submissions. Table~\ref{table_proposed} reports the results of the four acoustic features, which show that LFCC performs the best among them, followed by CQCC. It is observed that IFCC based phase features are not that effective for this classification, in fact perform poorer than MFCC features. This may be due to the fact that signal phase may not undergo very major shift in phase while wearing a mask to have effective classification only on the phase information. However, LFCC and CQCC emerge as strong acoustic feature based single systems that suggest the importance of linear filterbanks and long-term information. In addition, we note that LFCC outperforms all the four baselines of ComParE 2020 on the development set for classification of speech with and without masks.

We then perform the fusion studies with the acoustic features and the four ComParE 2020 baselines. Table~\ref{table_proposed} shows the score level fusion of the four acoustic features as well as the four baselines and their comparison to given challenge. We observe that the score level fusion of the baselines leads to an improved result than the challenge benchmark baseline using majority voting based fusion. This may be due to the fact that the majority voting considers fusion based on predicted class output of each system, whereas the score level fusion employs a weighted scheme to fuse various systems. We observe that the score level fusion of the acoustic features performs better than the individual feature that reveals complementary acoustic properties captured by each of them. Finally, the score level fusion of the acoustic features and the four baselines achieves an UAR of 73.50\% on the test set that beats the benchmark baseline by 1.70\%.

\section{Conclusions}
\label{conc}

This work devotes on studying novel acoustic features capturing different acoustic properties of a signal to classify speech with and without mask for ComParE 2020 challenge participation. We focused on features derived using linear filterbanks, instantaneous phase and long-term information of signal. Among these the linear filterbank and long-term information based features are found to be more effective for the challenge mask task. The systems of acoustic features are fused with the ComParE 2020 baselines at score level for challenge submission. The submitted system outperformed the challenge benchmark system based on majority voting of the best baselines showing usefulness of both score level fusion and acoustic feature based complementary information for classification of speech with and without mask.

\section{Acknowledgements}
This research work is supported by Programmatic Grant No. A1687b0033 from the Singapore Government's Research, Innovation and Enterprise 2020 plan (Advanced Manufacturing and Engineering domain) and also by Human-Robot Interaction Phase 1 (Grant No. 192 25 00054) by the National Research Foundation, Prime Minister's Office, Singapore under the National Robotics Programme.


\balance

\bibliographystyle{IEEEtran}

\bibliography{MyReferences_new}

\begin{thebibliography}{10}
\providecommand{\url}[1]{#1}
\csname url@samestyle\endcsname
\providecommand{\newblock}{\relax}
\providecommand{\bibinfo}[2]{#2}
\providecommand{\BIBentrySTDinterwordspacing}{\spaceskip=0pt\relax}
\providecommand{\BIBentryALTinterwordstretchfactor}{4}
\providecommand{\BIBentryALTinterwordspacing}{\spaceskip=\fontdimen2\font plus
\BIBentryALTinterwordstretchfactor\fontdimen3\font minus
  \fontdimen4\font\relax}
\providecommand{\BIBforeignlanguage}[2]{{%
\expandafter\ifx\csname l@#1\endcsname\relax
\typeout{** WARNING: IEEEtran.bst: No hyphenation pattern has been}%
\typeout{** loaded for the language `#1'. Using the pattern for}%
\typeout{** the default language instead.}%
\else
\language=\csname l@#1\endcsname
\fi
#2}}
\providecommand{\BIBdecl}{\relax}
\BIBdecl

\bibitem{HRI2012}
N.~C. Kr{\"a}mer, A.~von~der P{\"u}tten, and S.~Eimler, \emph{Human-Agent and
  Human-Robot Interaction Theory: Similarities to and Differences from
  Human-Human Interaction}.

\bibitem{Mask_maskless_2015}
A.~Nieto{-}Rodr{\'{\i}}guez, M.~Mucientes, and V.~M. Brea, ``Mask and maskless
  face classification system to detect breach protocols in the operating
  room,'' in \emph{ACM International Conference on Distributed Smart Camera
  2015}, 2015, pp. 207--208.

\bibitem{Schuller_ComParE2020}
B.~Schuller, A.~Batliner, C.~Bergler, E.-M. Messner, A.~Hamilton,
  S.~Amiriparian, A.~Baird, G.~Rizos, M.~Schmitt, L.~Stappen, H.~Baumeister,
  A.~D. MacIntyre, and S.~Hantke, ``The {INTERSPEECH} 2020 computational
  paralinguistics challenge: Elderly emotion, breathing \& masks,'' in
  \emph{Interspeech 2020}, 2020.

\bibitem{SARS_facemask}
D.~Coniam, ``The impact of wearing a face mask in a high-stakes oral
  examination: An exploratory post-{SARS} study in {Hong Kong},''
  \emph{Language Assessment Quarterly}, vol.~2, no.~4, pp. 235--261, 2005.

\bibitem{speaking_underCover}
N.~Fecher and D.~Watt, ``Speaking under cover: The effect of face-concealing
  garments on spectral properties of fricatives,'' in \emph{International
  Congress of Phonetic Sciences (ICPhS) 2011}, 2011, pp. 663--666.

\bibitem{surgical_mask_2008}
L.~L. Mendel, J.~A. Gardino, and S.~R. Atcherson, ``Speech understanding using
  surgical masks: A problem in health care?'' \emph{Journal of the American
  Academy of Audiology}, vol.~19, p. 686–695, 2008.

\bibitem{Mirco_surgeryRoom_2013}
M.~Ravanelli, A.~Sosi, M.~M.~M. Omologo, M.~Benetti, and G.~Pedrotti, ``Distant
  talking speech recognition in surgery room : The {DOMHOS} project,'' in
  \emph{AISV 2013}, 2013, p. 13 pages.

\bibitem{audio-visual_faceCover}
N.~Fecher, ``The `audio-visual face cover corpus': Investigations into
  audio-visual speech and speaker recognition when the speaker's face is
  occluded by facewear,'' in \emph{Interspeech 2012}, 2012, pp. 2250--2253.

\bibitem{facecover_is2015}
R.~Saeidi, T.~Niemi, H.~Karppelin, J.~Pohjalainen, T.~Kinnunen, and P.~Alku,
  ``Speaker recognition for speech under face cover,'' in \emph{Interspeech
  2015}, 2015, pp. 1012--1016.

\bibitem{Face_mask_is2016}
R.~Saeidi, I.~Huhtakallio, and P.~Alku, ``Analysis of face mask effect on
  speaker recognition,'' in \emph{Interspeech 2016}, 2016, pp. 1800--1804.

\bibitem{Schuller_IEEE_magazine}
B.~{Schuller}, ``The computational paralinguistics challenge,'' \emph{IEEE
  Signal Processing Magazine}, vol.~29, no.~4, pp. 97--101, July 2012.

\bibitem{SCHULLER2011speechComm}
B.~Schuller, A.~Batliner, S.~Steidl, and D.~Seppi, ``Recognising realistic
  emotions and affect in speech: State of the art and lessons learnt from the
  first challenge,'' \emph{Speech Communication}, vol.~53, no.~9, pp. 1062 --
  1087, 2011.

\bibitem{SCHULLER2013_para}
B.~Schuller, S.~Steidl, A.~Batliner, F.~Burkhardt, L.~Devillers, C.~M{\"u}ller,
  and S.~Narayanan, ``Paralinguistics in speech and language state-of-the-art
  and the challenge,'' \emph{Computer Speech \& Language}, vol.~27, no.~1, pp.
  4 -- 39, 2013.

\bibitem{SCHULLER2019_lessons}
B.~Schuller, F.~Weninger, Y.~Zhang, F.~Ringeval, A.~Batliner, S.~Steidl,
  F.~Eyben, E.~Marchi, A.~Vinciarelli, K.~Scherer, M.~Chetouani, and
  M.~Mortillaro, ``Affective and behavioural computing: Lessons learnt from the
  first computational paralinguistics challenge,'' \emph{Computer Speech \&
  Language}, vol.~53, pp. 156 -- 180, 2019.

\bibitem{ASVspoof2015_lfcc}
M.~Sahidullah, T.~Kinnunen, and C.~Hanilçi, ``A comparison of features for
  synthetic speech detection,'' in \emph{Proc. Interspeech 2015}, 2015, pp.
  2087--2091.

\bibitem{Vijayan_Speech_comm}
K.~Vijayan, P.~R. Reddy, and K.~S.~R. Murty, ``Significance of analytic phase
  of speech signals in speaker verification,'' \emph{Speech Communication},
  vol.~81, pp. 54 -- 71, 2016, phase-Aware Signal Processing in Speech
  Communication.

\bibitem{CQCC_odyssey2016}
M.~Todisco, H.~Delgado, and N.~Evans, ``A new feature for automatic speaker
  verification anti-spoofing: Constant {Q} cepstral coefficients,'' in
  \emph{Odyssey 2016}, 2016, pp. 283--290.

\bibitem{rkd_APSIPA2018}
R.~K. Das and H.~Li, ``Instantaneous phase and excitation source features for
  detection of replay attacks,'' in \emph{Asia-Pacific Signal and Information
  Processing Association Annual Summit and Conference (APSIPA ASC)}, Honolulu,
  Hawaii, 2018, pp. 1030--1037.

\bibitem{rkd_is2019_spoof}
R.~K. Das, J.~Yang, and H.~Li, ``Long range acoustic features for spoofed
  speech detection,'' in \emph{Interspeech 2019}, 2019, pp. 1058--1062.

\bibitem{rkd_is2019_orca}
R.~K. Das and H.~Li, ``Instantaneous phase and long-term acoustic cues for orca
  activity detection,'' in \emph{Interspeech 2019}, 2019, pp. 2418--2422.

\bibitem{rkd_ASRU2019}
R.~K. {Das}, J.~{Yang}, and H.~{Li}, ``Long range acoustic and deep features
  perspective on {ASVspoof} 2019,'' in \emph{IEEE Automatic Speech Recognition
  and Understanding Workshop (ASRU) 2019}, 2019, pp. 1018--1025.

\bibitem{Davis1980}
S.~Davis and P.~Mermelstein, ``Comparison of parametric representations for
  monosyllabic word recognition in continuously spoken sentences,'' \emph{IEEE
  Transactions on Acoustics, Speech and Signal Processing}, vol.~28, no.~4, pp.
  357--366, Aug 1980.

\bibitem{ComParE_functionals}
B.~Schuller, S.~Steidl, A.~Batliner, A.~Vinciarelli, K.~R. Scherer,
  F.~Ringeval, M.~Chetouani, F.~Weninger, F.~Eyben, E.~Marchi, M.~Mortillaro,
  H.~Salamin, A.~Polychroniou, F.~Valente, and S.~Kim, ``The {INTERSPEECH} 2013
  computational paralinguistics challenge: social signals, conflict, emotion,
  autism,'' in \emph{Interspeech 2013}, 2013, pp. 148--152.

\bibitem{openXBoW}
M.~Schmitt and B.~Schuller, ``{OpenXBOW}: Introducing the passau open-source
  crossmodal bag-of-words toolkit,'' \emph{Journal of Machine Learning
  Research}, vol.~18, no.~1, pp. 3370--3374, Jan. 2017.

\bibitem{DeepSpectrum}
S.~Amiriparian, M.~Gerczuk, S.~Ottl, N.~Cummins, M.~Freitag, S.~Pugachevskiy,
  A.~Baird, and B.~Schuller, ``Snore sound classification using image-based
  deep spectrum features,'' in \emph{Interspeech 2017}, 2017, pp. 3512--3516.

\bibitem{auDeep}
M.~Freitag, S.~Amiriparian, S.~Pugachevskiy, N.~Cummins, and B.~Schuller,
  ``{auDeep}: Unsupervised learning of representations from audio with deep
  recurrent neural networks,'' \emph{Journal of Machine Learning Research},
  vol.~18, no.~1, pp. 1--5, 2018.

\bibitem{audeep_fused}
S.~{Arniriparian}, M.~{Freitag}, N.~{Cummins}, M.~{Gerczuk}, S.~{Pugachevskiy},
  and B.~{Schuller}, ``A fusion of deep convolutional generative adversarial
  networks and sequence to sequence autoencoders for acoustic scene
  classification,'' in \emph{26th European Signal Processing Conference
  (EUSIPCO)}, Sep. 2018, pp. 977--981.

\bibitem{RevModPhys}
H.~Fletcher, ``Auditory patterns,'' \emph{Reviews of Modern Physics}, vol.~12,
  pp. 47--65, 1940.

\bibitem{tifs_subband}
J.~Yang, R.~K. Das, and H.~Li, ``Significance of subband features for synthetic
  speech decetion,'' \emph{IEEE Transactions on Information Forensics and
  Security}, vol.~15, pp. 2160--2170, 2020.

\bibitem{Marple1997}
S.~L. Marple, ``Computing the discrete-time `analytic' signal via {FFT},'' in
  \emph{Conference Record of the Thirty-First Asilomar Conference on Signals,
  Systems and Computers}, vol.~2, Nov 1997, pp. 1322--1325.

\bibitem{ASVspoof2017_Jelil2017}
S.~Jelil, R.~K. Das, S.~R.~M. Prasanna, and R.~Sinha, ``Spoof detection using
  source, instantaneous frequency and cepstral features,'' in \emph{Proc.
  Interspeech 2017}, 2017, pp. 22--26.

\bibitem{Brown1991}
J.~C. Brown, ``Calculation of a constant {Q} spectral transform,''
  \emph{Journal of Acoustical Society of America}, vol.~89, pp. 425--434, 1991.

\bibitem{CQCC_CSL}
M.~Todisco, H.~Delgado, and N.~Evans, ``Constant {Q} cepstral coefficients: {A}
  spoofing countermeasure for automatic speaker verification,'' \emph{Computer
  Speech {\&} Language}, vol.~45, pp. 516--535, 2017.

\bibitem{pyknogram_ref}
A.~Potamianos and P.~Maragos, ``Speech formant frequency and bandwidth tracking
  using multiband energy demodulation,'' \emph{The Journal of the Acoustical
  Society of America}, vol.~99, no.~6, pp. 3795--3806, 1996.

\bibitem{Opensmile1}
F.~Eyben, M.~W\"{o}llmer, and B.~Schuller, ``{OpenSmile}: The munich versatile
  and fast open-source audio feature extractor,'' in \emph{ACM International
  Conference on Multimedia}, 2010, pp. 1459--1462.

\bibitem{Opensmile2}
F.~Eyben, F.~Weninger, F.~Gross, and B.~Schuller, ``Recent developments in
  {openSMILE}, the munich open-source multimedia feature extractor,'' in
  \emph{Proceedings of the 21st ACM International Conference on
  Multimedia}.\hskip 1em plus 0.5em minus 0.4em\relax ACM, 2013, pp. 835--838.

\bibitem{cvpr16}
K.~He, X.~Zhang, S.~Ren, and J.~Sun, ``Deep residual learning for image
  recognition,'' in \emph{{IEEE} Conference on Computer Vision and Pattern
  Recognition (CVPR) 2016}, 2016, pp. 770--778.

\bibitem{Reynolds1995}
D.~A. Reynolds and R.~Rose, ``Robust text-independent speaker identification
  using {Gaussian} mixture speaker models,'' \emph{IEEE Transactions on Speech
  and Audio Processing}, vol.~3, no.~1, pp. 72--83, Jan 1995.

\bibitem{rkd_is}
R.~K. Das, {Abhiram B.}, S.~R.~M. Prasanna, and A.~G. Ramakrishnan, ``Combining
  source and system information for limited data speaker verification,'' in
  \emph{Interspeech 2014}, 2014, pp. 1836--1840.

\bibitem{rkd_ncc_2015}
R.~K. Das, D.~Pati, and S.~R.~M. Prasanna, ``Different aspects of source
  information for limited data speaker verification,'' in \emph{National
  Conference on Communications (NCC) 2015}, 2015, pp. 1--6.

\bibitem{rkd_jasa}
R.~K. Das and S.~R.~M. Prasanna, ``Exploring different attributes of source
  information for speaker verification with limited test data,'' \emph{The
  Journal of the Acoustical Society of America}, vol. 140, no.~1, pp. 184--190,
  2016.

\bibitem{Bosaris}
N.~Br{\"{u}}mmer and E.~de~Villiers, ``The {BOSARIS} toolkit: Theory,
  algorithms and code for surviving the new {DCF},'' \emph{CoRR}, vol.
  abs/1304.2865, 2013.

\end{thebibliography}

\end{document}